\documentclass[pre,aps,floatfix,authordate1-4]{revtex4-1}

\usepackage{rotating} 
\usepackage{times}
\usepackage{graphicx}
\usepackage{setspace}
\usepackage{amsmath}
\usepackage[multiple]{footmisc}
\usepackage[normalem]{ulem}

\begin{document}
%\singlespacing

\title{Towards atomistic resolution structure of phosphatidylcholine glycerol backbone and choline headgroup at different ambient conditions}

\author{Alexandru Botan}
\thanks{The authors are listed in alphaphetical order.}
%\thanks{The author list is not completed.}
\thanks{Institut Lumi\`ere Mati\`ere, UMR5306 Universit\'e Lyon 1-CNRS, Universit\'e de Lyon 69622 Villeurbanne, France}
\author{Andrea Catte}
\thanks{University of East Anglia, Norwich, United Kingdom}
\author{Fernando Favela}
\author{Patrick Fuchs}
\thanks{Institut Jacques Monod, CNRS, Université Paris Diderot, Sorbonne Paris Cité, Paris, France}
%\thanks{Université Paris Diderot, Sorbonne Paris Cité, Paris, France}
\author{Matti Javanainen}
\thanks{Tampere University of Technology, Tampere, Finland}
\author{Waldemar Kulig}
\thanks{Tampere University of Technology, Tampere, Finland}
\author{Antti Lamberg}
\thanks{Department of Chemical Engineering, Kyoto University, Kyoto, Japan}
%\author{Alexander Lyubartsev}
\author{Markus S. Miettinen}
\thanks{Fachbereich Physik, Freie Universität Berlin, Berlin, Germany}
\author{Luca Monticelli}
\thanks{5IBCP, CNRS UMR 5086, Lyon, France}
\author{Jukka M\"a\"att\"a}
\thanks{Aalto University, Espoo, Finland}
\author{Vasily S. Oganesyan}
\thanks{University of East Anglia, Norwich, United Kingdom}
\author{O. H. Samuli Ollila} 
\thanks{{\bf Author to whom correspondence may be addressed. E-mail: samuli.ollila@aalto.fi.}}
\thanks{Helsinki Biophysics and Biomembrane Group, Department of Biomedical Engineering and Computational Science, Aalto University, Espoo, Finland}
\author{Marius Retegan}
\thanks{Max Planck Institute for Chemical Energy Conversion, Mülheim an der Ruhr, Germany}
\author{Hubert Santuz}
\thanks{INSERM, U1134, DSIMB; Institut National de la Transfusion Sanguine (INTS); Laboratoire d'Excellence GR-Ex, Paris, France}
\thanks{Université Paris Diderot, Sorbonne Paris Cité, Paris, France}
\author{Joona Tynkkynen}
\thanks{Tampere University of Technology, Tampere, Finland}
%\author{Mark Wilson}
%\author{Alexander Vogel}

\begin{abstract}
Phospholipids are essential building blocks of biological membranes.
Despite of vast amount of accurate experimental data the atomistic resolution structures sampled by the glycerol backbone and choline headgroup
in phoshatidylcholine bilayers are not known. Atomistic resolution molecular dynamics simulation model 
would automatically resolve the structures giving an interpretation of experimental results, if the model
would reproduce the experimental data. In this work we compare the C-H bond vector order
parameters for glycerol backbone and choline headgroup between 14 different atomistic resolution
models and experiments in fully hydrated lipid bilayer. The current models are not accurately enough to resolve the structure.
However, closer inspection of three best performing models (CHARMM36, GAFFlipid and MacRog) suggest
that improvements in the sampled dihedral angle distributions would potentilly lead to the model which
would resolve the structure. Despite of the inaccuracy in the fully hydrated
structures, the response to the dehydration, i.e. P-N vector tilting more parallel to membrane normal, 
is qualitatively correct in all models. The CHARMM36 and MacRog models describe the interactions
between lipids and cholesterol better than Berger/H\"oltje model.
This work has been, and continues to be, progressed and discussed through the blog: nmrlipids.blogspot.fi. 
Everyone is invited to join the discussion and make contributions through the blog. 
The manuscript will be eventually submitted to an appropriate scientific journal. 
Everyone who has contributed to the work through the blog will be offered 
coauthorship. For more details see: nmrlipids.blogspot.fi.

\end{abstract}

\maketitle

~\vspace{0.3cm}\\
{\it \bf} 

\section{Introduction}

\begin{figure}[]
  \centering
  \includegraphics[width=8.6cm]{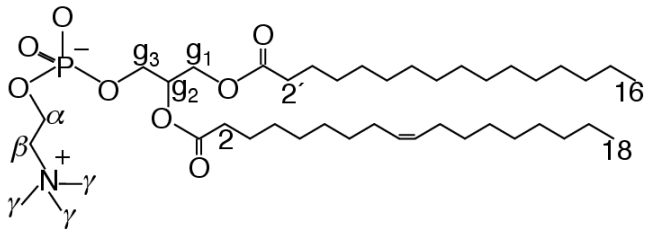}

  \caption{\label{POPCstructure}
    Chemical structure of  1-palmitoyl-2-oleoylphosphatidylcholine (POPC).}
  
\end{figure}

Phospholipids containing different polar headgroups and different acyl chains are essential building blocks of 
biological membranes. Lamellar phospholipid bilayer structures have been widely studied with various experimental 
and theoretical techniques as a simple model for the biological membranes~\cite{lipowsky95,tieleman97,klauda08,edholm08,tieleman10,piggot12,rabinovich13,marsh13}. 
Phospholipid molecules are composed of hydrophobic acyl chains and hydrophilic headgroup, which are connected by glycerol backbone,
see Fig.~\ref{POPCstructure} for the structure of 1-palmitoyl-2-oleoylphosphatidylcholine (POPC).
The behaviour of the acyl chains in a bilayer is relatively well understood~\cite{Israelachvili80,lipowsky95,tieleman97,klauda08,edholm08,tieleman10,marsh13}. 
The structures sampled by the glycerol backbone and choline in liquid bilayer state, however, are not fully 
resolved since even the most accurate scattering and Nuclear Magnetic Resonance (NMR)
techniques give only a set of values that the structure has to fulfill, but
there is no unique way to derive the actual structure from these parameters~\cite{seelig77b,skarjune79,Israelachvili80,jacobs80,davis83,strenk85,akutsu91,hong95b,hong96,semchyschyn04}.
Some structural details have been extracted from crystal structure, $^1$H NMR studies and Raman spectroscopy~\cite{hauser80,hauser81,hauser81b,akutsu81b,pascher92,hauser88,marsh06}
but general consensus about the structures sampled in the liquid state have not been reached~\cite{seelig77b,skarjune79,Israelachvili80,jacobs80,davis83,strenk85,hauser88,akutsu91,hong95b,hong96,semchyschyn04,marsh06}. 
On the other hand, the glycerol backbone structures are similar for various biologically
relevant lipid species (phosphatidylcholine (PC), phosphatidylethanolamine (PE) and phosphatidylglycerol (PG)) 
in various environments~\cite{gally81} and the headgroup choline structures are similar in model membranes and
real cells (mouse fibroblast L-M cell)~\cite{scherer87}.
Thus, the resolvation of phosphatidylcholine glycerol backbone and choline structures would be 
useful for understanding wide range of different biological membranes.

Classical atomistic resolution molecular dynamics simulations have been widely used to study  
lipid bilayers~\cite{tieleman97,klauda08,edholm08,tieleman10,piggot12,rabinovich13}. As these models provide an atomistic
resolution description of the whole lipid molecule, they have potential to resolve the glycerol backbone and 
headgroup structures. In particular, the experimental C-H bond order parameters for the glycerol backbone 
(g$_1$, g$_2$ and g$_3$) and choline ($\alpha$ and $\beta$) segments (see Fig.~\ref{POPCstructure} for definitions) are among the main parameters used in
attempts to derive the structures from experimental data~\cite{seelig77b,skarjune79,jacobs80,davis83,akutsu91,hong95b,semchyschyn04}.
These parameters are also routinely compared between experiments and simulations for the acyl chains~\cite{tieleman97,klauda08,edholm08,tieleman10,piggot12}.
Thus, the structures sampled in a simulation model that reproduces these and other experimental parameters, automatically
give an interpretation of the experiments, in other words they can be considered as reasonable atomistic resolution descriptions of
the behaviour of lipid molecules in a bilayer.

The glycerol backbone and choline headgroup order parameters have been compared between simulations and experiments
in some studies~\cite{shinoda97,hogberg08,castro08,klauda10,kapla12,dickson12,poger12,ferreira13,chowdhary13,maciejewski14}, 
however much less frequently than for acyl tail chains~\cite{tieleman97,klauda08,edholm08,tieleman10,piggot12}.
The main reason is probably that the existing experimental data for the glycerol backbone
and choline headgroups is scattered over many publications and published in a format that is difficult to understand without some NMR expertise. 
In addition to the order parameters, also dihedral angles for glycerol backbone and headgroup estimated from experiments have been sometimes used to 
assess the quality of simulation model~\cite{robinson94,essex94,kothekar96,hyvonen97,shinoda97,duong99}.

In this work we first review the most relevant experimental data for the glycerol backbone and choline headgroup order parameters
in a phosphatidylcholine lipid bilayer. Then the available atomistic resolution lipid models are carefully compared to the 
experimental data. The comparison reveals that the CHARMM36~\cite{klauda10}, GAFFlipid~\cite{dickson12} and MacRog models~\cite{maciejewski14}
have the most realistic glycerol backbone and choline structures. We also compare the glycerol backbone and choline 
structures between the most often used (Berger based) lipid model~\cite{berger97} 
%\todo{Based on the results and discussions with Olle Edholm and Peter Tieleman in the blog, it seems that 
%the glycerol backbone dihedrals are not the same in the model used in the publication by Berger et al.~\cite{berger97}
%and in the model shared in the webpage of Peter Tieleman (http://wcm.ucalgary.ca/tieleman/downloads).
%In this work (as it is quite often done in other works as well) we currently refer to the model in the webpage as ''Berger'' model.
%This is misleading. We have to call it something else. One suggestion could be to call it ''BergerM'', because it was firt time
%used in Marrink et al. Biophys. J. 74,931-943 (1998). Any other suggestions?\\
%COMMENT: Patrick Fuchs suggested ''Berger-T''. This is better. \\
%COMMENT2: Current trial, based in blog discussion, is to use Berger-MOLECULE-YEAR [citation].
%}
and the best performing models, to demonstrate that by using the 
order parameters we can distinguish the more reasonable structures from the less reasonable ones. However, none of the current models 
is accurate enough to resolve the atomistic resolution stuctures.

In addition to the fully--hydrated single--component lipid bilayers, the glycerol backbone and choline order parameters
have been measured under a large number of different conditions. For example, as a function of hydration level~\cite{bechinger91,ulrich94,dvinskikh05b}, cholesterol content~\cite{brown78,ferreira13}
ion concentration~\cite{brown77,akutsu81,altenbach84,roux90,roux91}, temperature~\cite{gally75}, charged lipid content~\cite{roux90,roux91}, charged surfactant content~\cite{scherer89}, 
drug molecule concentration~\cite{browning82,kelusky84,castro08}, and protein content~\cite{roux89,kuchinka89} (listing only the publications most relevant for this work and the pioneering studies).
Awareness of the existence of this type of data allows the comparison of structural responses to varying conditions between simulations and experiments,
which can be used to validate the simulation models and to interpret the original experiments. 
In this publication we demonstrate the power of this approach for understanding the behaviour of a bilayer as a function of hydration level and cholesterol concentration.
Choline headgroup order parameters as function of ion concentration, and their relation to the ion binding affinity, are discussed elsewhere~\cite{ionpaper}.

\section{methods}

\subsection{Open collaboration}
%\todo{Should we write more about this?}

This work has been done as an open collaboration by using nmrlipids.blogspot.fi as an communication platform.
The approach is inspired by the Polymath project~\cite{gowers09}, however there are some essential differences. The manuscript pointing out problems in the glycerol backbone and
headgroup structure in the most used molecular dynamics simulation model for lipid bilayers (Berger based models) was used as a starting point~\cite{ollila13}.
After publishing the initial manuscript, the blog was opened for further contributions and discussion from anyone interested. 
All the contributions were done publicly through the blog and the blog contributors were offered coauthorship. The final author list
is based on the self-assesment of the authors scientific contribution to the project.

Large portion of the used simulation files and scripts are shared through the GitHub organization https://github.com/NMRlipids
and Zenodo community https://zenodo.org/collection/user-nmrlipids.

\subsection{Analysis}
%\todo{ We should mention the sign issue and refer to the blog post:
%http://nmrlipids.blogspot.fi/2014/04/on-signs-of-order-parameters.html somewhere.
%Maybe we should change the name of this section II.B. to: "Analysis and terminology of order parameters from experiments and simulations" or similar, and also move the "forking" explanation here.}
The order parameter of a hydrocarbon C-H vector is defined as $S_{{\rm CH}}=\frac{3}{2}\langle \cos^2 \theta-1 \rangle$, where
the average is an ensemble average over the sampled conformations, and $\theta$ is the angle between the C-H bond and the membrane normal.
The order parameters can be measured by detecting quadrupolar splitting with $^2$H NMR~\cite{seelig77c} or by detecting dipolar 
splitting with $^{13}$C NMR~\cite{hong95a,gross97,dvinskikh05a,ferreira13}. The measurements are based on
different physical interactions and also the connection between order parameters and quadrupolar or dipolar splitting
are different. The order parameters from the measured quadrupolar splitting $\Delta \nu_Q$ ($^2$H NMR) are calculated using 
the equation $|S_{{\rm CD}}|=\frac{4}{3} \frac{e^2qQ}{h} \Delta \nu_Q$, where the value for the static quadrupole
splitting constant is estimated from various experiments to be 170~kHz leading to a numerical relation $|S_{{\rm CD}}|=0.00784 \times \Delta \nu_Q$~\cite{seelig77c}. 
The order parameters from the measured dipolar splitting $d_\mathrm{CH}$ ($^{13}$C NMR) are calculated using equation
$|S_{{\rm CH}}|=\frac{4\pi\langle r_\mathrm{CH}^3 \rangle}{\hbar \mu_0 \gamma_h \gamma_c} d_\mathrm{CH}$, where
values between 20.2-22.7 kHz are used for $\frac{4\pi\langle r_\mathrm{CH}^3 \rangle}{\hbar \mu_0 \gamma_h \gamma_c}$,
depending on the original authors~\cite{hong95a,gross97,dvinskikh05a,ferreira13}.
It is important to note that the order parameters measured with different techniques based on different physical interactions are in good agreement
with each other (see Results and Discussion), indicating very high quantitative accuracy of the measurements.
For more detailed discussion see http://nmrlipids.blogspot.fi/2014/02/accuracy-of-order-\-parameter-measurements.html
%\todo{This is temporarily linked to the blog. Currently I think that the posts (with comments included) to which we want to cite are put into 
%the Figshare in pdf format as they are in when the publication is submitted. Figshare allows commenting to continue the discussion.}.

The order parameters from simulations were calculated directly using the definition.
%$S_{{\rm CH}}=\frac{3}{2} \langle \cos^2 \theta-1 \rangle$. 
For the united atom models the hydrogen locations were generated 
post-simulationally using the positions of the heavy atoms in the simulation trajectories.
The statistical error estimates were calculated for the best performing simulation
models by calculating the error of the mean for the average over individual lipids in the system. 

It has been recently pointed out that the sampling of individual dihedral angles might be very
slow compared to the typical simulation timescales~\cite{vogel12}. On the other hand, another recent
study shows that the slowest rotational correlation functions of C-H bond (g$_1$) reaches plateau (S$_{CH}^2$)
after 200ns in the Berger-POPC-07~\cite{ollila07a} model, and that the dynamics of this segment is significantly too slow in simulations
compared to the experiments~\cite{ferreira15}. In practise, less than 200~ns data set is enough for the order parameter
calculation due to the average over different lipid molecules. In conclusion, if the sampling with typical simulation times
is not enough for the convergense of the order parameters, then the simulation models has significantly too slow dynamics.

\subsection{Simulated systems}
All simulations are ran with a standard setup for planar lipid bilayer in zero tension and constant temperature
with periodic boundary conditions in all directions by using Gromacs software package~\cite{hess08} 
(version numbers 4.5-4.6) or LAMMPS~\cite{plimpton95}.
The number of molecules, temperatures and the length of simulations for all the simulated systems 
are listed in Tables~\ref{systems},~\ref{systemsDEHYD} and~\ref{systemsCHOL}. Full simulation
details are given in the Supplementary Information (SI) or in the original publications if the
data is used previously. For some systems also the simulation related files and trajectories are publicly
available. The references pointing to simulation details and files are also listed in Tables~\ref{systems},~\ref{systemsDEHYD} and~\ref{systemsCHOL}.

\begin{table*}[htb]
\centering
\caption{Simulated single component lipid bilayers with full hydration. The simulation file data sets marked with $^*$ include also part of the trajectory.
  If simulation data from previously published work has been directly used, the original publication is cited for simulation details. For other systems the simulation details
  are given in Supplementary Information. 
}\label{systems}
\begin{tabular}{c c c c c c c c c}
%\hline
Force field & lipid  & \footnote{The number of lipid molecules}N$_{\rm l}$   &  \footnote{The number of water molecules}N$_{\rm w}$ & \footnote{Simulation temperature}T (K)  & \footnote{The total simulation time}t$_{{\rm sim}}$(ns) & \footnote{Time frames used in the analysis}t$_{{\rm anal}}$ (ns) & \footnote{Reference link for the downloadable simulation files}Files  &  \footnote{Reference for the full simulation details} Details\\
\hline
Berger-POPC-07~\cite{ollila07a}          &   POPC & 128 & 7290  & 298  & 270 & 240 & \cite{bergerFILESpopc}$^*$ & \cite{ferreira15} \\
%\sout {Berger-POPC-07~\cite{ollila07a}}          &   POPC & 128 & 7290  & 323  & 56 & 56  & ? & SI \\
Berger-DPPC-98~\cite{marrink98}          &   DPPC & 72 & 2864  & 323  & 140 & 100  & \cite{bergerDPPCfiles} & SI \\
Berger-DMPC-04~\cite{gurtovenko04}          &   DMPC & 128 & 5097  & 323  & 130 & 100  & \cite{dmpcFILES} & \cite{miettinen09} \\
CHARMM36\cite{klauda10}       & POPC   & 72  &  2242 & 303 & 30 & 20  & \cite{charmm36filesSHORT}$^*$ & SI \\
CHARMM36\cite{klauda10}    & POPC   & 128 &  5120    & 303 & 150 & 100  & - & SI \\
CHARMM36\cite{klauda10}       & DPPC   & 72  &  2189 & 323 & 30 & 25  & - & SI \\
%CHARMM36\cite{klauda10}       & DOPC   & 128 &  5120    & 303 & 150 & 100  & ? & SI \\
MacRog\cite{maciejewski14}  & POPC & 288  & 12600 & 310 & 100 & 80  & \cite{macrogFILES}$^*$ & SI  \\
GAFFlipid\cite{dickson12}       & POPC & 126  & 3948  & 303 & 137 & 32  & \cite{GAFFlipidFILES}$^*$ & SI \\
Lipid14\cite{dickson14}         & POPC  & 72 & 2234 & 303 & 100 & 50  & \cite{lipid14files}$^*$ & SI \\
Poger\cite{poger10}             & DPPC  & 128 & 5841 & 323 & 2$\times$100 & 2$\times$50 & \cite{pogerFILES} & SI \\
Slipid\cite{jambeck12}          & DPPC & 128 & 3840 & 323 & 150 & 100 & \cite{slipidsFILES}$^*$ & SI \\
Kukol\cite{kukol09}          & POPC   & 512 & 20564 & 298 & 50 & 30  & \cite{kukolFILES}$^*$ & SI \\
Chiu et al.\cite{chiu09}     & POPC  & 128 & 3552  & 298 & 56 & 50  & - & SI \\
Rabinovich et al.\cite{rabinovich14}  & POPC   &  128 & 3840  & 303 & 100 & 80  & - & \cite{rabinovich14}  \\
H\"ogberg et al.\cite{hogberg08}  & DMPC   &  98 & 3840  & 303 & 75 & 50 & - & \cite{hogberg08} \\
Ulmschneider\cite{Ulmschneider09}    & POPC  & 128 & 3328 & 310 & 100 & 50 & \cite{ulmschneiderFILES}$^*$ & SI \\
Tj\"ornhammar et al.\cite{tjornhammar14}   & DPPC  & 144 & 7056 & 323 & 200 & 100 & \cite{tjornhammarfiles}$^*$ & \cite{tjornhammar14} \\
%\sout{OPLS-AA}\cite{rog09b}\todoi{We have this only with 150mM of NaCl delivered by Joona Tynkkynen. Options are to remove these results from this publicatio, run the simulations without ions (if not yet available), 
%  or include the results with ions. In my understanding this is a proto version of MacRog so it could be left out as well. However, for historical reasons and 
%to understand literature it might be useful to include also these results. COMMENT: Matti Javanainen suggested that we would remove these results. If there are no objections and the
%data without ions will not be delivered, we will remove this.}         &   &  &  & & & & & \\
CHARMM36-UA~\cite{henin08,lee14}     & DLPC   & 128  & 3840  & 323 & 50 & 20 & \cite{charmmUAfiles} & SI \\
\end{tabular}
\end{table*} 

\begin{table*}[htb]
\centering
\caption{Simulated single component lipid bilayers with varying hydration levels. The simulation file data sets marked with $^*$ include also part of the trajectory.
}\label{systemsDEHYD}
\begin{tabular}{c c c c c c c c c c}
%\hline
Force field & lipid & \footnote{Water/lipid molar ratio} n (w/l)   & \footnote{The number of lipid molecules}N$_{\rm l}$   &  \footnote{The number of water molecules}N$_{\rm w}$ & \footnote{Simulation temperature}T (K)  & \footnote{The total simulation time}t$_{{\rm sim}}$(ns)  & \footnote{Time frames used in the analysis}t$_{{\rm anal}}$ (ns)& \footnote{Reference link for the downloadable simulation files}Files  &  \footnote{Reference for the full simulation details} Details\\
\hline
Berger-POPC-07~\cite{ollila07a}          &   POPC & 57  &128 & 7290  & 298  & 270 & 240 & \cite{bergerFILESpopc}$^*$ & SI \\
                                        &   POPC & 7  &128 & 896   & 298  & 60 & 50 & \cite{bergerDEHYDfiles}$^*$ & SI \\
CHARMM36\cite{klauda10}              & POPC   & 31 & 72  &  2242 & 303 & 30 & 20 & \cite{charmm36filesSHORT}$^*$ & SI \\
                               & POPC   & 15 & 72 &  1080  & 303 & 59 & 40 & \cite{charmm36files15wPERl}$^*$ & SI \\
                            & POPC   & 7  & 72  &  504  & 303 & 60 & 20 & \cite{charmm36files7wPERl}$^*$ & SI \\
MacRog\cite{maciejewski14}     & POPC   & 50 & 288  & 14400 & 310 & 90 & 40 & \cite{macrogdehydFILES}$^*$ & SI \\
                               & POPC   & 25 & 288  & 7200 & 310 & 100 & 50 & \cite{macrogdehydFILES}$^*$ & SI \\
                                & POPC   & 20 & 288  & 5760 & 310 & 100 & 50 & \cite{macrogdehydFILES}$^*$ & SI \\
                                & POPC   & 15 & 288  & 4320 & 310 & 100 & 50 & \cite{macrogdehydFILES}$^*$ & SI \\
                                & POPC   & 10 & 288  & 2880 & 310 & 100 & 50 & \cite{macrogdehydFILES}$^*$ & SI \\
                                & POPC   & 5 & 288   & 1440 & 310 & 100 & 50 & \cite{macrogdehydFILES}$^*$ & SI \\
GAFFlipid\cite{dickson12}      & POPC   & 31& 126  & 3948  & 303 & 137 & 32 & \cite{GAFFlipidFILES}$^*$ & SI  \\
                               & POPC   & 7& 126  & 896   & 303 & 130 & 40 & \cite{gaffDEHYDfiles}$^*$ & SI \\
\end{tabular}
\end{table*} 

\begin{table*}[htb]
\centering
\caption{Simulated lipid bilayers with cholesterol. The simulation file data sets marked with $^*$ include also part of the trajectory.
}\label{systemsCHOL}
\begin{tabular}{c c c c c c c c c c}
%\hline
Force field & lipid  & \footnote{The number of lipid molecules}N$_{\rm l}$  & \footnote{number of cholesterol molecules}N$_{\rm chol}$   &  \footnote{The number of water molecules}N$_{\rm w}$ & \footnote{Simulation temperature}T (K)  & \footnote{The total simulation time}t$_{{\rm sim}}$(ns)  & \footnote{Time frames used in the analysis}t$_{{\rm anal}}$ (ns)\\
\hline
Berger-POPC-07~\cite{ollila07a}/H\"oltje-CHOL-13~\cite{holtje01,ferreira13}          &   POPC &128 & 0 & 7290  & 298  & 270 & 240 & \cite{bergerFILESpopc}$^*$ & \cite{ferreira15} \\
                               &   POPC &120 & 8 & 7290   & 298  & 100 & 80 & \cite{bergerFILESpopc7chol}$^*$ & \cite{ferreira13} \\
                               &   POPC &110 & 18 & 8481  & 298  & 100 & 80 & \cite{bergerFILESpopc15chol}$^*$ & \cite{ferreira13}  \\
                               &   POPC &84 & 44 & 6794   & 298  & 100 & 80 & \cite{bergerFILESpopc34chol}$^*$ & \cite{ferreira13} \\
                               &   POPC &64 & 64 & 10314  & 298  & 100 & 80 & \cite{bergerFILESpopc50chol}$^*$ & \cite{ferreira13} \\
                               &   POPC &50 & 78 & 5782   & 298  & 100 & 80 & \cite{bergerFILESpopc60chol}$^*$ & \cite{ferreira13} \\
CHARMM36\cite{klauda10,lim12}    & POPC   & 128&  0 & 5120  & 303  & 150 & 100 & - & SI  \\
                            & POPC   & 100  & 24  &  4960   & 303 & 200 & 100 & - & SI \\
                    & POPC   & 80  & 80  &  4496    & 303 & 200 & 100 & - & SI \\
MacRog\cite{maciejewski14}     & POPC   & 128 & 0  & 6400   & 310 & 400 & 200 & \cite{macrogCHOLfiles}$^*$ & SI \\ 
                          & POPC   & 114  & 14 & 6400  & 310  & 400 & 200 & \cite{macrogCHOLfiles}$^*$ & SI    \\
                          & POPC   & 72   & 56 & 6400  & 310  & 400 & 200 & \cite{macrogCHOLfiles}$^*$ & SI    \\
                             & POPC   & 64  & 64 & 6400  & 310  & 400 & 200 & \cite{macrogCHOLfiles}$^*$ & SI    \\
                             & POPC   & 56   & 72 & 6400  & 310  & 400 & 200 & \cite{macrogCHOLfiles}$^*$ & SI    \\
\end{tabular}
\end{table*} 

%The initial structures for the simulations in low hydrated conditions are made by removing the
%water molecules from the fully hydrated system to achieve the targeted hydration conditions.

%For the initial structure of CHARMM36-~\cite{klauda10,lim12} and MacRog-~\cite{maciejewski14,kulig14} simulations with cholesterol,
%the required amount of lipid molecules were replaced with cholesterol 

\section{Results and Discussion}

\subsection{Full hydration: Experimental order parameters for glycerol backbone and headgroup}\label{experiments}
\begin{figure}[]
  \centering
  \includegraphics[width=8.6cm]{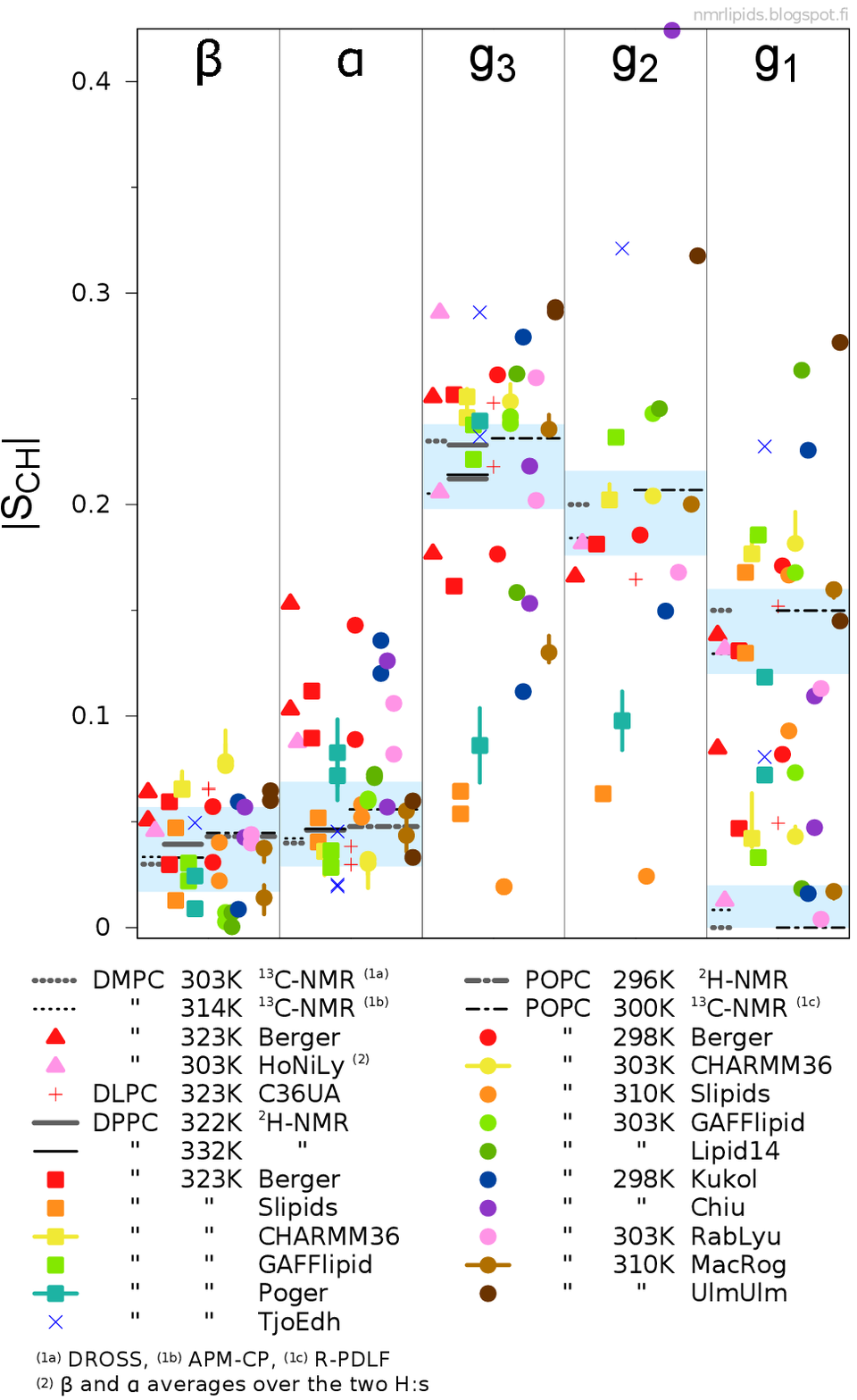}
%\todo{DONE}\\
%The data in this figure should be updated to be consistent with the systems listed in Table~\ref{systems}. At least, Tjornhammar et al and CHARMM-UA values are missing. Some results in the figure are outdated due to longer simulations reported etc.} \\
%todo{Markus Miettinen suggested that we should consider making one more figure where only experimental data would be shown and that would be discussed in Section~\ref{experiments}} \\
%todo{There has been some discussion in the blog that maybe we should include the signs and stereospecific labeling here as well, and that maybe this figure should be clarified. My view is that for the main conclusion, i.e. CHARMM36, GAFFlipid and MacRog are the best here, this format is good enough. However, if someone wants to improve this and/or collect all the data with the signs and/or stereospecific labeling it would be even better. There is also very interesting discussion about interactive figure.}
  \caption{\label{HGorderparameters}
  Order parameteres from simulations listed in Table~\ref{systems} and experiments for glycerol and choline groups.
  The experimental values were taken from the following publications: DMPC 303K from \cite{gross97}, DMPC 314K from \cite{dvinskikh05a}, DPPC 310K from \cite{gally75}, 
  DPPC 323K from \cite{akutsu81}, and POPC 298K \cite{ferreira13}.
  The vertical bars shown for some of the computational values are not error bars, but demonstrate that for 
  these systems we had two data sets; the ends of the bars mark the values from the two sets, and the dot marks their measurement-time-weighted average. 
  The H{\"o}gberg et al. force field in Table~\ref{systems} is abbreviated in this figure as HoNiLy, Tj{\"o}rnhammar et al. as TjoEdh, CHARMM36-UA as C36UA,
  Rabinovic et al. as RabLyu and Ulmschneider as UlmUlm.
} 
\end{figure}

The specific deuteration of $\alpha$-, $\beta$- and g$_3$- segments of dipalmitoylphosphatidylcholine (DPPC) has been successful, 
allowing the order parameter measurements for these segments by $^2$H--NMR~\cite{gally75,brown77,brown78,akutsu81}.
In addition, the order parameters for all glycerol backbone and choline headgroup segments in egg yolk lecithin~\cite{hong95a},
1,2-dimyristoyl-sn-glycero-3-phosphocholine (DMPC)~\cite{hong95b,gross97,dvinskikh05a}, 
1,2-dioleoyl-sn-glycero-3-phosphocholine (DOPC)~\cite{warschawski05} and POPC~\cite{warschawski05,ferreira13}
have been measured with several different implementations of $^{13}$C NMR experiments.
The experimental absolute values of glycerol backbone and choline order parameters from various publications are shown in Fig.~\ref{HGorderparameters}.

In general there is a good agreement between the order parameters measured with different experimental NMR techniques: Almost all the 
reported values are inside variation of $\pm$0.02 (which is also the error estimate given by Gross et al.~\cite{gross97}) 
for all fully hydrated PC bilayer, regardless of the variation in their acyl chain composition and the temperature.
Exception are the somewhat lower order parameters sometimes reported having been measured with $^{13}$C--NMR~\cite{hong95a,hong95b,warschawski05}.
These experiments have not seen in Fig.~\ref{HGorderparameters} as the reported error bars are either relatively large~\cite{hong95a,hong95b}, 
or the spectral resolution is quite low and the numerical lineshape simulations have not been used in the analysis~\cite{warschawski05}.
Therefore it is highly likely that these reported lower order parameters are due to lower experimental 
accuracy and that we exclude these values from our discussion. 
Motivated by the high experimental repeatability, we have highlighted in 
Fig.~\ref{HGorderparameters} subjective sweet spots (light blue areas), within which we expect the calculated absolute 
values of order parameters of a well-performing force field to fall.

%In addition to the phosphatidylcholine lipids, similar values of the glycerol backbone order parameters have been measured
%for phophatidylethanolamine (PE) and phopshatidylglycerol (PG) in E. Coli extract~\cite{gally81},
%indicating that the glycerol backbone structure is similar independent of the headgroup chemistry and lipid environment.
%Further, choline order parameters measured from mouse fibroblast L-M cell are similar to the ones in model
%membranes~\cite{scherer87}.

In addition to the numerical values, an important feature of the glycerol backbone is the 
inequality of order parameters for the two hydrogens attached to the same carbon in g$_1$ and g$_3$ segments,
while the hydrogens in choline $\alpha$ and $\beta$ segments give equal values.
Note that in this work we call the phenomena of inequal order parameters for hydrogens attached to the same carbon as ''forking'' to avoid 
confusion with dipolar and quadrupolar splitting in NMR terminology. Forking is also observed experimentally for the C$_2$ carbon in the sn-2
chain of all phosholipids, and it is known to arise from differently sampled orientations of the two C-H bonds, not from two 
different populations of lipid conformations~\cite{engel81}. The forking in glycerol backbone g$_3$ segment is small ($\approx$ 0.02) 
and some experiments only report the larger or the average value~\cite{akutsu81,ferreira13}. 
In contrast, forking is significant for the glycerol backbone g$_1$ segment, whose lower order parameter is close to zero and the
larger one has absolute values around 0.13-0.15. Forking was studied in detail by Gally et al.~\cite{gally81}, who used E. Coli to 
stereospecifically deuterate the different hydrogens attached to the g$_1$ or g$_3$ groups in PE lipids, and measured the order parameters from the lipid 
extract. This experiment gave the lower order parameter when deuterium was in the S position of g$_1$ or R position for g$_3$.
Since the glycerol backbone order parameters are very similar irrespective of the headgroup chemistry (PC,PE and PG) or lipid 
environment~\cite{gally81}, it is reasonable to assume that the stereospecifity measured for the PE lipids
holds also for the PC lipids.

In Fig.~\ref{HGorderparameters} we have shown the absolute values of order parameters as these are accessible
with both $^2$H NMR and $^{13}$C NMR techniques. However, $^{13}$C NMR techniques allow also the measurement of 
the sign of the order parameter~\cite{hong95a,hong95b,gross97}. The measured sign is negative for almost all the carbons 
discussed in this work, only $\alpha$ is positive~\cite{hong95a,hong95b,gross97}. 

\begin{figure*}[]
  \centering
  \includegraphics[width=8.6cm]{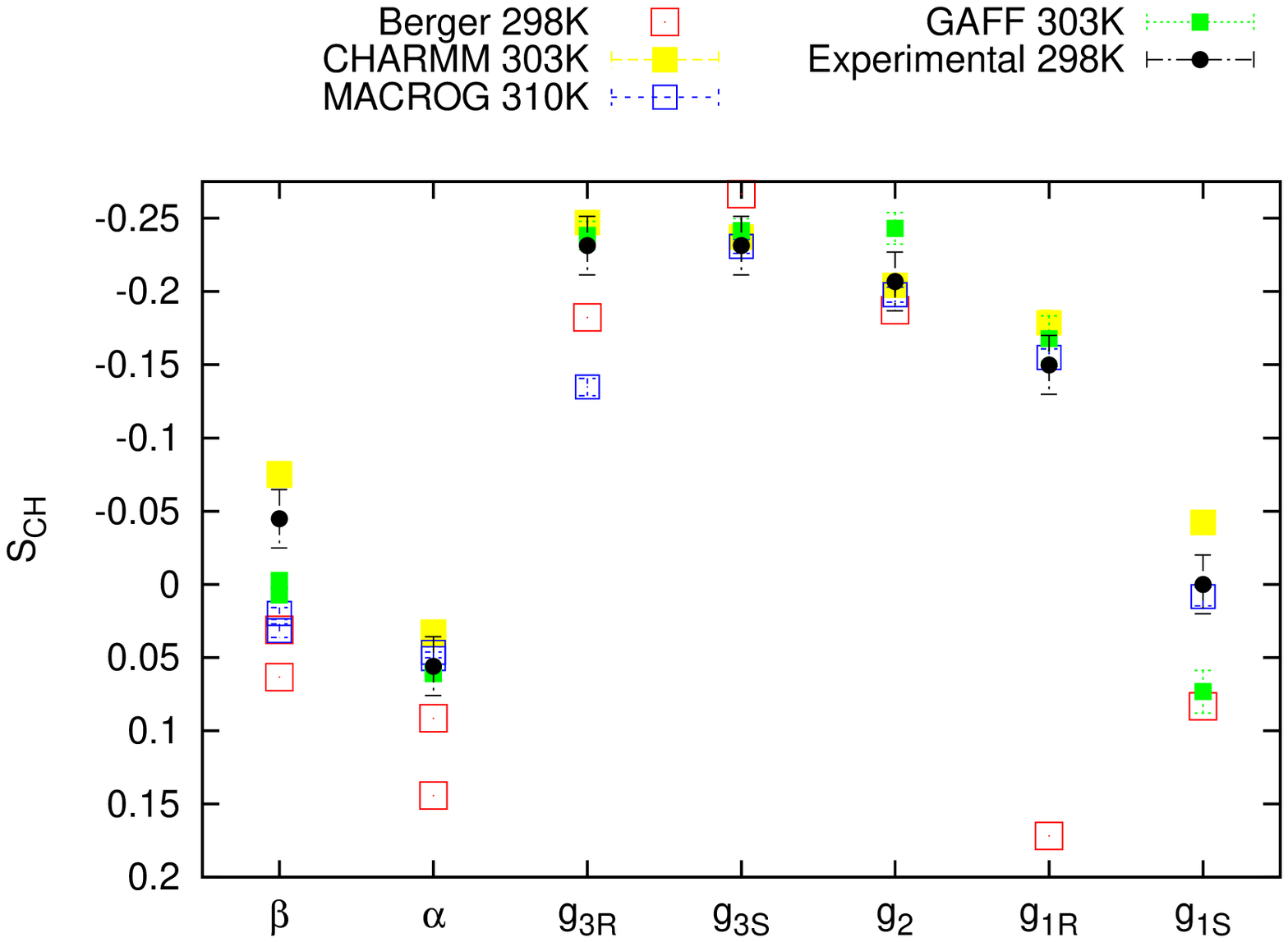} \\
  %todo{Markus Miettinen suggested that we should consider ``spreading'' the values similarly to the figure~\ref{HGorderparameters}} \\
  \caption{\label{HGorderparameters2}
  Order parameteres from simulations with Berger-POPC-07, CHARMM36, GAFFlipid and Macrog force fields together with experimental values for POPC glycerol and choline groups.
  The magnitudes for experimental order parameters are taken from Ferreira et al.~\cite{ferreira13}, the signs are based on the measurements by Hong et al.~\cite{hong95a,hong95b} 
  and Gross et al.~\cite{gross97}, and the R/S labeling is based in the measurements by Gally et al.~\cite{gally81}.
} 
\end{figure*}

Combining the experimental information of the sign~\cite{hong95a,hong95b,gross97} and the stereospecifity 
measurements~\cite{gally81} with the absolute value measurements from various techniques~\cite{gally75,akutsu81,gross97,dvinskikh05a,ferreira13}
having high quantitative accuracy,
the most detailed experimentally available order parameter information for the glycerol backbone and choline segments of POPC is obtained.
These data are shown in Fig.~\ref{HGorderparameters2}.

\subsection{Full hydration: Comparison between simulation models and experiments}

The order parameters of the glycerol backbone and headgroup calculated from different force fields for various lipids has been 
previously compared to experiments~\cite{shinoda97,hogberg08,castro08,klauda10,kapla12,dickson12,poger12,ferreira13,chowdhary13,maciejewski14}. 
The general conclusion from these works seems to be that the CHARMM based~\cite{hogberg08,klauda10}, GAFFlipid~\cite{dickson12} and
MacRog~\cite{maciejewski14} force fields perform better for the glycerol backbone and headgroup structures than the Gromos based models~\cite{castro08,kapla12,poger12,ferreira13}.
However, none of the studies exploits the full potential of the available experimental data discussed in previous section; the quantitative accuracy, known signs and stereospecific labeling of
the experimental order parameters.

To get a general idea about the quality of the glycerol backbone and choline headgroup structures in different models, we calculated the absolute
values of the order parameters for these parts from fourteen different lipid models (Table~\ref{systems}) and 
plotted the results together with experimental values in Fig.~\ref{HGorderparameters}.
Two criteria were used to judge the quality of the model: {\bf forkings} there must not be significant forking in the $\alpha$ and $\beta$ carbons,
there must be only moderate forking in the g$_3$ carbon and there must be significant forking in the g$_1$ carbon, {\bf absolute values}
should be preferably inside to the subjective sweet spots determined from experiments (blue shaded regions in Fig.~\ref{HGorderparameters}).
None of the studied force fields fulfills these criteria, however, three force fields are closer than others: CHARMM36~\cite{klauda10}, MacRog~\cite{maciejewski14} and GAFFlipid~\cite{dickson12}.
%he results for each force field in respect to the above criteria are summarized in Table~\ref{??} \todo{This kind of table should be done or not?}.

The top three models (CHARMM36, MacRog and GAFFlipid) together with the most used lipid model (Berger based model) 
were subjected to a more careful comparison including the signs and the sterospecific labeling in Fig.~\ref{HGorderparameters2}.
The essential additional information given by this comparison is that the sign of the $\beta$ carbon order parameter is correct only in CHARMM36 model.

\subsection{Full hydration: Atomistic resolution structures in different models}

The results in the previous section revealed significant differences of the glycerol backbone and choline headgroup
order parameters between different molecular dynamics simulation models.
However, it is not straightforward to conclude which kind of structural differences (if any)
between the models the results indicate, because the mapping from the order parameters to the 
structure is not unique. In this section we demonstrate that 1) the differences in order parameters
indicate significantly different structural sampling strongly correlating with the dihedral angles of the related bonds,
and that 2) the comparison between experimental and simulated order parameters can be used to exclude
nonrealistic structural samping in molecular dynamics simulations. The demonstration is done for 
the dihedral angles defined by the g$_3$-g$_2$-g$_1$-O(sn-1) segments in the glycerol backbone and 
the N-$\beta$-$\alpha$-O segments in the headgroup. These dihedrals were chosen for demonstration, because 
significant differences between the models are observed around these segments in Fig.~\ref{HGorderparameters2}.
We note that performing a similar comparison through all the dihedrals in all the 14 models would probably give highly useful
information to improve the accuracy of simulations, however this is beyond the scope of the current report. 

\begin{figure}[!t]
\vspace{0.3cm}
  \centering
  \includegraphics[width=8.6cm]{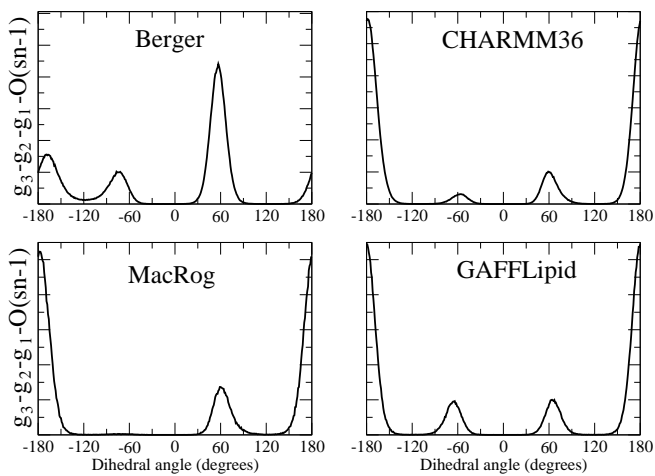}
\vspace{0.3cm}
  \caption{\label{dihDISTS}
    Dihedral angle distributions for g$_3$-g$_2$-g$_1$-O(sn-1) dihedral from different models (POPC bilayer in full hydration).
      } 
\end{figure}

The dihedral angle distributions for the  g$_3$-g$_2$-g$_1$-O(sn-1) dihedral calculated from different models are
shown in Fig.~\ref{dihDISTS}. The distribution is qualitatively different for the Berger-POPC-07 model, showing a maximum in 
the gaughe$^+$-conformation (60$^o$) compared to all the other models showing a maximum in the trans-conformation (180$^o$).
The distributions in all the other models have the same general features, the main difference being that the
fraction of configurations in gaughe$^-$-conformation (-60$^o$) is zero for the MacRog, detectable for the CHARMM36 and
equally large to the gaughe$^+$ fraction in GAFFlipid. From the results we conclude that most likely the wrongly sampled
dihedral angle for the g$_2$-g$_1$ bond explains the significant discrepancy to the experimental order parameters
for the g$_1$ segment in the Berger-POPC-07 model (Fig.~\ref{HGorderparameters2}). 
The result that models preferring the trans conformation for this dihedral gives more realistic order parameters
is in agreement with previous crystal structure and $^1$H NMR studies~\cite{hauser80,hauser81,hauser81b,hauser88,pascher92,marsh06}.

\begin{figure}[]
  \centering
  \includegraphics[width=8.6cm]{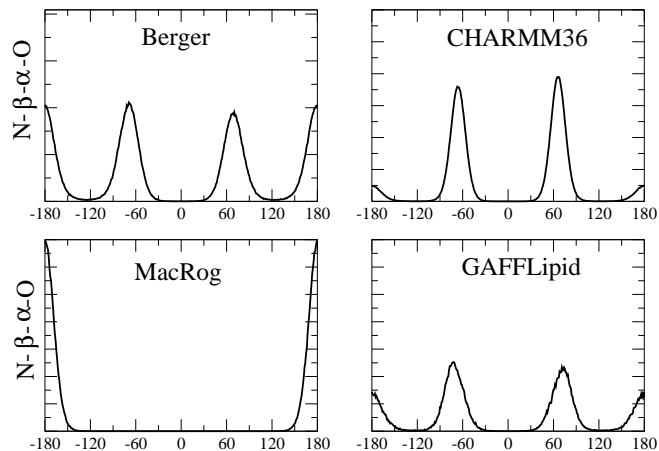}
\vspace{0.5cm}
  \caption{\label{dihDISTS2}
    Dihedral angle distributions for N-$\beta$-$\alpha$-O dihedral from different models (POPC bilayer in full hydration).
  } 
\end{figure}

The dihedral angle distribution for the  N-$\beta$-$\alpha$-O dihedral calculated from the same four models is 
shown in Fig.~\ref{dihDISTS2}. Also for this dihedral there are significant differences in the gauche-trans fractions.
The gaughe conformations are dominant in the CHARMM36, in MacRog there are only trans conformations present,
and in the Berger-POPC-07 and GAFFlipid gaughe and trans conformation has equal probability. 
On the other hand comparison of $\alpha$ and $\beta$ order parameters in Fig.~\ref{HGorderparameters2}
reveals that the CHARMM36 is closest to the experimental results being the only model having correct
sign (negative) for the $\beta$ order parameter. This result is again in agreement with previous 
crystal structure, $^1$H NMR and Raman spectroscopy studies~\cite{hauser80,hauser81,hauser81b,akutsu81b} which suggest that
this dihedral has only gaughe conformation in the absense of ions.

%Interestingly, the probability of the gaughe conformations correlates with the order parameter difference between the $\beta$ and $\alpha$ segments:
%the larger the gaughe fraction the larger the order parameter difference. This suggestion together with the results in 
%Fig.~\ref{HGorderparameters2} would indicate that the correct gaughe-trans fraction for N-$\beta$-$\alpha$-O dihedral is 
%larger than in GAFFlipid but smaller than in CHARMM36. This happens to be quite close to the gaughe-trans fraction in
%the Berger model. However, there is significant forking and numerical values are off from the experiments in the Berger
%model suggesting that it has some other inaccuracies.

The used examples show that the glycerol backbone and headgroup order parameters reflect the atomistic resolution structure
and that the comparison with experiments allows the assesment of the quality of the suggested structure. We were able to pinpoint
specific problems in the structures in different models and suggest potential improvement strategies.
If the improved atomistic resolution
molecular dynamics simulation model would reproduce the order parameters and other experimental observables (like chemical shift anisotropy)
with experimental accuracy, it would give an interpretation for the atomistic resolution structure of the glycerol backbone and 
choline~\cite{seelig77b,skarjune79,jacobs80,davis83,akutsu91,hong95b,semchyschyn04}. The research along these lines is left, however,
for future studies.

\subsection{Response to dehydration and cholesterol content}
In addition to pure phosphatidylcholine bilayers at full hydration, the choline headgroup order parameters
have been measured under various different conditions~\cite{gally75,brown77,brown78,akutsu81,altenbach84,scherer89,bechinger91,ulrich94,dvinskikh05b,castro08,kapla12,ferreira13}.
Also the order parameters for the glycerol backbone have been measured with $^{13}$C NMR in dehydrated conditions~\cite{dvinskikh05b}, and as a function 
of anesthetics~\cite{castro08} and glycolipids~\cite{kapla12} for DMPC, and as a function of cholesterol 
concentration for POPC~\cite{ferreira13}. Due to the high resolution in the NMR (especially $^2$H NMR) experiments,
even very small order parameter changes resulting from the varying conditions can be measured (see
http://nmrlipids.blogspot.fi/2014/02/accuracy-of-order-parameter-measurements.html
for more discussion.
%todo{This is temporarily linked to the blog. Currently I think that the posts (with comments included) to which we want to cite are put into 
%he Figshare in pdf format as they are in when the publication is submitted. Figshare allows commenting to continue the discussion.}
However, as already discussed above, it is not simple to deduce 
the structural changes from order parameter changes~\cite{akutsu91,semchyschyn04}. Consequently, comparison of the order parameters
between simulations and experiments in different conditions can be used measure the quality of the force field 
in different situations, and, if the quality is good, to potentially interpret the structural changes in experiments.
Here we exemplify such comparison for a lipid bilayer under low hydration levels and mixed with cholesterol. 
The interaction between ions and phosphatidylcholine bilayer is discussed in a separate work~\cite{ionpaper}.

\subsubsection{Phospholipid bilayer with low hydration level}
\begin{figure}[]
  \centering
  \includegraphics[width=8.6cm]{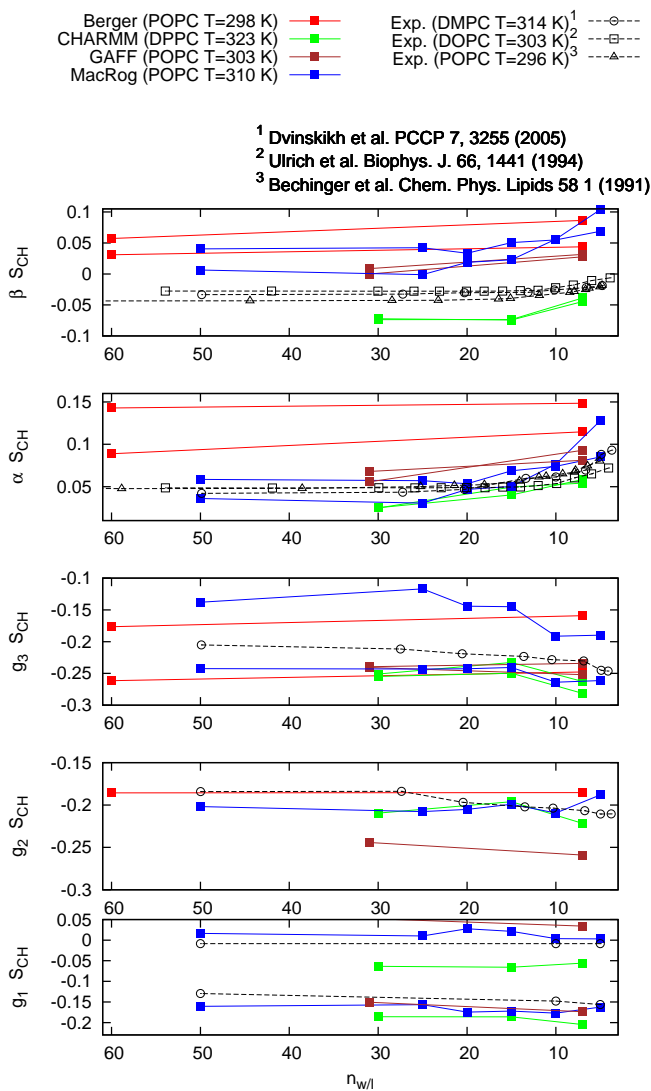}
%todo{Several people think that this plot should be clarified.}
  \caption{\label{ordPhydr}
    The effect of dehydration on glycerol and choline order parameters in experiments.
    The magnitudes of order parameters are measured for DMPC ($^{13}$C NMR) at 314~K~\cite{dvinskikh05b}, 
    for POPC ($^2$H NMR) at 296~K~\cite{bechinger91} and for DOPC ($^2$H NMR) at 303~K~\cite{ulrich94}. 
    The signs are based on the measurements by Hong et al.~\cite{hong95a,hong95b} 
    and Gross et al.~\cite{gross97}.
  }
\end{figure}

The experimental order parameters available in the literature~\cite{dvinskikh05b,ulrich94,bechinger91} 
for the glycerol backbone and choline as a function of hydration level are shown in Fig.~\ref{ordPhydr}. 
The independently reported values for choline segments are in good agreement with each other (despite of 
slight differences in temperature and acyl chain composition),
showing increase for both segments with decreasing hydration level. It should be noted that only 
absolute values were measured in the original experiments~\cite{dvinskikh05b,ulrich94,bechinger91}, but
we have included the signs measured separately~\cite{hong95a,hong95b,gross97}. 
Consequently, the $\beta$ order parameter with negative sign actually increases with dehydration 
since the absolute value decreases~\cite{dvinskikh05b,ulrich94,bechinger91}.
Slight decrease for the glycerol backbone g$_3$- and g$_2$- order parameters were observed with dehydration, 
while g$_1$ remained practically unchanged~\cite{dvinskikh05b}.

Lipid bilayer dehydration has been studied also with molecular dynamics simulations~\cite{mashl01,pertsin05,pertsin07,eun09,eun10,schneck12},
typically motivated by the  discussion about the origin of the ``hydration repulsion''~\cite{israelachvili,israelachvili96,sparr11}.
However, the used simulation models are not typically compared to the experimental choline and glycerol backbone
order parameters (except by Mashl et al.~\cite{mashl01}).
In Fig.~\ref{ordPhydr} the glycerol backbone and choline order parameters as a function of hydration level are shown 
for the CHARMM36, MacRog and GAFFlipid models (having the most realistic atomistic resolution structures) together with the Berger based model 
(which is the most used lipid model). The choline order parameters increase with dehydration in all simulation
models, in qualitative agreement with experiments. 
The measured decrease in both g$_3$ and g$_2$ order parameters with dehydration is reproduced only in CHARMM36.

\begin{figure}[]
  \centering
  \includegraphics[width=8.6cm]{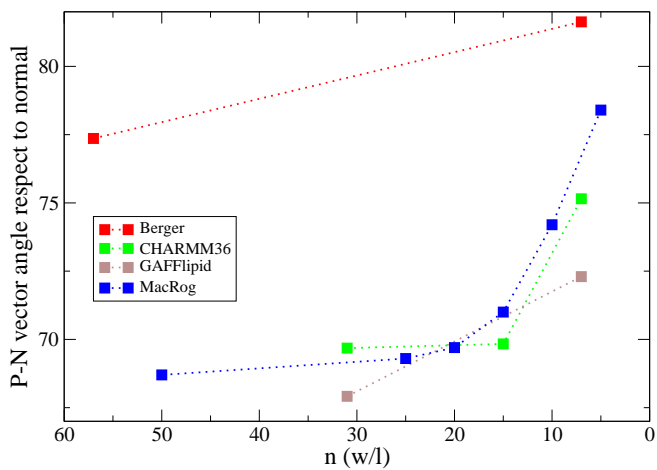}

  \caption{\label{PNangle}
    The angle between membrane normal and P-N vector in choline segment as function of
    hydration level calculated from different simulations.
  }
\end{figure}

The qualitative agreement with experiments in all simulation models for the $\alpha$ and $\beta$ order parameters  
as a function of hydration indicates that the structural response of the choline headgroup to dehydration is somewhat realistic
despite the unrealistic structures at full hydration. 
The most likely explanation is that the choline group
orients more parallel to the membrane plane with dehydration due to restricted interlamellar space. 
Indeed, the P-N angle vector angle with membrane normal as a function of dehydration shows an increase for
all models as a function of dehydration in Fig.~\ref{PNangle}.
However, the qualitative agreement in the lipid response to dehydration does not guarantee the correct 
free energy landscape if the simulation model has incorrect structure. The influence of this issue to 
dehydration energetics studied with simulations~\cite{eun09,schneck12} is left for future studies.

The response of the glycerol backbone to dehydration seems to be more subtle than of the choline headgroup 
as CHARMM36 is the only model that reproduces the decrease in g$_2$ and g$_3$ segments.

\subsubsection{Phospholipid bilayer mixed with cholesterol}
Phospholipid--cholesterol interactions have been widely studied with theoretical~\cite{huang99,zhu07,rog09,alwarawrah12} and
experimental methods~\cite{brown78,marsh10,ferreira13,marsh13}, since cholesterol is abundant in biological membranes and
it has been suggested to be an important player, for example, in domain formation~\cite{simons04,somerharju09}.
It is widely agreed that cholesterol orders lipid acyl tails thus decreasing the area per molecule (condensing effect),
however, the influence of cholesterol on the lipid headgroup and glycerol backbone are sill debated~\cite{huang99,simons04,somerharju09}.
For example, it has been suggested that the surrounding lipids shield cholesterol from interactions with water by 
reorienting their headgroups (``umbrella model'')~\cite{huang99} or that cholesterol acts as a spacer for the headgroups thus increasing 
their entropy and dynamics (``superlattice model'')~\cite{somerharju09}. 
%\todo{It has been suggested that we should use direct quotations from original papers here for clarity.}
Both of these suggestions have been supported
by molecular dynamics simulations~\cite{zhu07,alwarawrah12}, and other simulations suggest specific
interactions between the glycerol backbone and cholesterol~\cite{rog09}, however the glycerol backbone and choline headgroup behaviour
as a function of cholesterol content is not compared to experiments in these studies. 

\begin{figure}[]
  \centering
  \includegraphics[width=8.6cm]{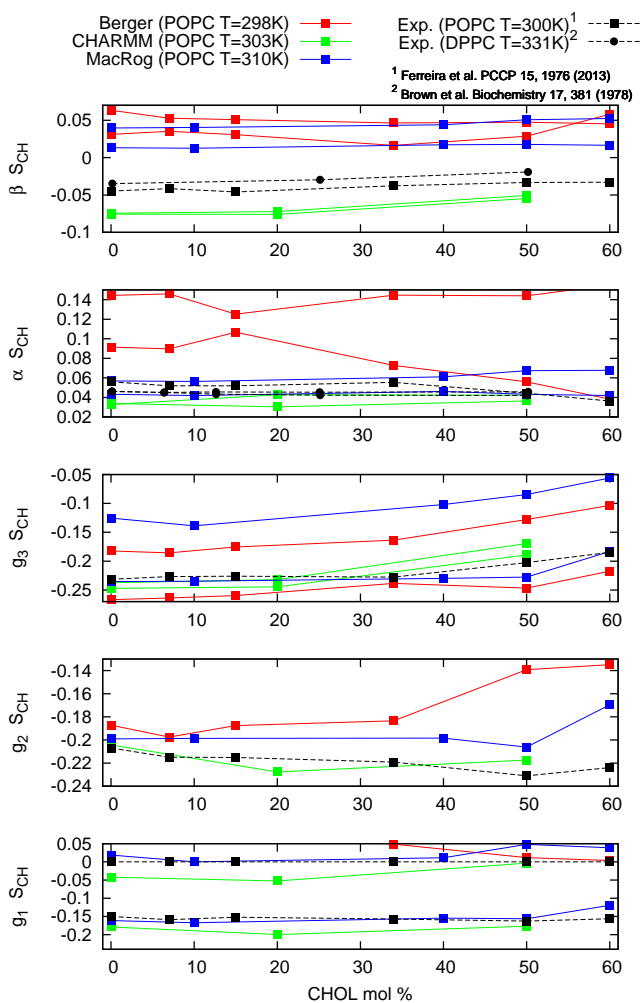}
  \caption{\label{ordPchol}
    The effect of cholesterol content on the glycerol backbone and choline order parameters in experiments~\cite{brown78,ferreira13} and simulations
    with the Berger-POPC-07/H\"oltje-CHOL-13, CHARMM36 and MacRog force fields. The signs in the experimental values are based on the measurements by Hong et al.~\cite{hong95a,hong95b} 
    and Gross et al.~\cite{gross97}.  The most order parameters from Berger-POPC-07/H\"oltje-CHOL-13 model for g$_1$ are beoynd the y-axis scale.}
\end{figure}

The choline headgroup and glycerol backbone order parameters for POPC measured by $^{13}$C NMR~\cite{ferreira13} and DPPC choline order parameters 
measured by $^{2}$H NMR~\cite{brown78} are shown in Fig.~\ref{ordPchol} as a function of cholesterol content.
The agreement between different experimental results is again very good, showing only very modest changes in 
the choline order parameters as a function of cholesterol content. It should be noted, however, that very small
changes are measurable with high resolution $^{2}$H NMR experiments
and cholesterol causes a measurable increase in the $\beta$ order parameter and a forking in the $\alpha$ order
parameter~\cite{brown78}, but these effects are so small that they are barely visible in the scale used in Fig.~\ref{ordPchol}.
Further, the effects of cholesterol on the glycerol backbone order parameters for POPC from $^{13}$C NMR experiment~\cite{ferreira13} 
is in good agreement with the results for the phosphatidylethanolamine (PE) measured by $^{2}$H NMR~\cite{ghosh82}.
These results further support the idea that the glycerol backbone structural behaviour is independent of the
headgroup composition~\cite{gally81} and that the headgroup stucture is independent of the acyl chain region content unless
charges are present~\cite{scherer87}.

In addition to the experimental data, the previously published simulation results from the Berger-POPC-07/H\"oltje-CHOL-13 model~\cite{ferreira13},
and our results from CHARMM36 and MacRog force fields  
are shown in Fig.~\ref{ordPchol}. As already pointed out previosly, the Berger-POPC-07/H\"oltje-CHOL-13 model
seriously overestimates the effect of cholesterol on the phospholipid glycerol backbone and choline segments~\cite{ferreira13}.
In contrast, the responses of both CHARMM36 and MacRog are in better agreement
with experiments, however CHARMM36 seems to better reproduce the experimentally observed modest changes in the glycerol backbone segments 
g$_2$ and g$_3$ with high concetrations of cholesterol. Thus we have calculated the glycerol backbone dihedral angle distributions
as a function of cholesterol in CHARMM36 (shown in Supplementary material) to resolve the cholesterol induced structural changes. The only detectable change due to the
addition of cholesterol is the small decrease of gaughe- and increase of gaughe+ probability of g3-g2-g1-O(sn-1) dihedral.

It should be noted that the CHARMM36 force field parameters (dihedral potentials) for the glycerol backbone have been tuned 
the dihedral potentials to reproduce the correct order parameters at fully hydrated conditions~\cite{klauda10}. 
This procedure contains a risk of overfitting, which would manifest itself as wrong responses to changing conditions. 
According to our results, tuning seems not to lead to overfitting problems in the case of dehydration or lipid-cholesterol mixtures.

\section{Conclusions}
The atomistic resolution structures sampled by the glycerol backbone and choline headgroup
in phoshatidylcholine bilayers are not known despite of vast amount of accurate experimental
data. Atomistic resolution molecular dynamics simulation model which would reproduce the
experimental data would automatically resolve the structures giving an interpretation of experimental results.
In this work we have collected and reviewed the experimental C-H bond vector order
parameters available in literature. These experimental parameters are then compared to
different atomistic resolution simulation models for fully hydrated bilayer, dehydrated bilayer and
lipid bilayer containing cholesterol. Our results have led to the following conclusions:

- The C-H bond order parameters measured with different NMR techniques are in good agreement
with each others. By combining the experimental results from various sources we concluded
that the order parameters for each C-H bond are known with quantitative accuracy of $\pm$0.02.

- None of the tested models (14 different models) produces the order parameters with the experimental
accuracy for fully hydrated phoshatidylcholine lipid bilayer. However, the CHARMM36, GAFFlipid and MacRog 
models are relatively close. The structures of these models together with the most used lipid model (Berger based) 
were subjected to more careful studies. The results revealed that the current models are not accurate
enough to resolve the atomistic resolution structures sampled by glycerol backbone and choline headgroup.  
However, the correlation between dihedral angle distributions and order parameter differences was found, 
suggesting that careful adjustment of dihedral potentials would potentially lead to the model with correct
structure.

- Independent of the accuracy for fully hydrated lipid bilayer, all the models reproduced the choline response
to the dehydration. This can be explained by the change in the P-N vector tilting more parallel to the membrane
which leads to the increase of order parameters despite of the initial configuration. It should be however noted
that the correct qualitative response do not necessarily indicate correct energetics. 

- The response of glycerol backbone and choline headgroup to the cholesterol content is described more
realistically in CHARMM36 and MacRog models than in the Berger based model.

In general, we conclude that the atomistic resolution classical molecular dynamics simulations 
is extremely convenient tool to give structural interpretation for the high resolution NMR data~\cite{ferreira14}. 
However, in the case of phoshatidylcholine glycerol backbone and choline headgroup there is some
further model development required.

This work has been, and continues to be, progressed and discussed through the blog: nmrlipids.blogspot.fi. 
Everyone is invited to join the discussion and make contributions through the blog. 
The manuscript will be eventually submitted to an appropriate scientific journal. 
Everyone who has contributed to the work through the blog will be offered 
coauthorship. For more details see: nmrlipids.blogspot.fi.

{\bf Acknowledgements: }
OHSO acknowledges Tiago Ferreira and Paavo Kinnunen for useful discussions, the Emil Aaltonen foundation for financial support, Aalto Science-IT project and CSC-IT Center for Science for computational resources. 

\bibliographystyle{apsrev}
\bibliography{refs}

\newpage

\appendix
\begin{center}
{\bf SUPPLEMENTARY INFORMATION}
\end{center}
\subsection{Simulation details} 
%\todo{When I was adding the simulation details to the manuscript, I realized that the section was getting very long and
%it would do the paper more difficult to read. I suggest that we do this in the following way: \\
%- For the simulation data which are taken directly from previous publications, we only cite the original publication
%in the tables where the systems are described. This would be the case, for example, for Berger POPC/chol, DMPC, Rabinovich et al. and Hogberg et al. \\
%- For the simulations which are reproduced using mdp file already delivered by original authors we can cite that, thus we do not have to copy the 
%simulation detail section from the original publication. This is the case, for example, for Tjornhammar et al. \\
%- When simulations are done specifically for this project and there is something non-trivial we
%put the details in the supplementary material. This is the case, for example, for GAFFlipid, Lipid14 and poger et al. \\
%- In all cases all simulation files, also trajectory, would be ideally shared in Zenodo.
%}
\subsubsection{Berger based models}
For the berger based models we use here the following naming convention: 
Berger - \{{\it molecule name}\} - \{{\it year when model published first time}\} \{{\it citation}\}.
The reason is that there are several different molecular topologies which are using the non-bonded parameters originally
developed by Berger et al.~\cite{berger97}. Thus the common factor in the berger based models are the non-bonded parameters,
while the molecule specific parameters might somewhat vary. However, the majority of the molecular level topologies are 
relying (especially for the glycerol backbone and headgroup) on the parameters originally introduced by Marrink et al.~\cite{marrink98}.
This is the case for all the Berger based simulations discussed in this work.

POPC simulations at full hydration in 298 K and simulations as a function of cholesterol are the same as previous publications~\cite{ferreira13,ferreira15}.
In these simulation the POPC parameters introduced by Ollila et al~\cite{ollila07a} are used, which are using Berger non-bonded parameters~\cite{berger97}
and molecular topology is from Tieleman et al.~\cite{tieleman99} with improved double bond dihedrals by Bachar et al.~\cite{bachar04}. 
Thus they are called Berger-POPC-07~\cite{ollila07a}. The cholesterol model is based on the parameters by H\"oltje et al.~\cite{holtje01} with the
exeption that the atom types were changed from CH2/CH3 to LP2/LP3 to avoid overcondensation of the bilayer as suggested in ref.~\cite{tieleman06}.
Since this modification was introduced by Ferreira et al.~\cite{ferreira13}, we call the used cholesterol model as H\"oltje-CHOL-13~\cite{ferreira13}.

For the POPC at 323~K and POPC in low hydration the same force field parameters are used.
For DPPC the implementation of Berger parameters~\cite{berger97} by Peter Tieleman et al. are used~\cite{marrink98}.
For all of these simulations the timestep of 2~fs was used with leap-frog integrator. Covalent bond lengths were constrained with LINCS algorithm~\cite{hess97,hess07}. 
Coordinates were written every 10~ps. PME~\cite{darden93,essman95} with real space cut-off at 1.0~nm was used 
for electrostatics. Plain cut-off was used for the Lennard-Jones interactions with a 1.0~nm cut-off.
The neighbour lists were updated every 5th step with cut-off at 1.0~nm. Temperature was coupled separately
for lipids and water to 298~K with the velocity-rescale method~\cite{bussi07} with coupling constant 0.1~ps.
Pressure was semi-isotropically coupled to the atmospheric pressure with the Berendsen method~\cite{berendsen84}.

%The results for the POPC/cholesterol systems with the Berger lipid~\cite{berger97} and modified H\"oltje
%cholesterol model~\cite{holtje01} were taken directly from Ref.~\cite{ferreira13}. 

\subsubsection{CHARMM36}

{\it DPPC}
%\todo{Markus Miettinen made the files.}

Timestep of 1~fs was used with leap-frog integrator. Covalent bonds with hydrogens were constrained with LINCS algorithm~\cite{hess97,hess07}. 
Coordinates were written every 5~ps. PME~\cite{darden93,essman95} with real space cut-off at 1.4~nm was used 
for electrostatics. Lennard-Jones interactions were switched to zero between 0.8~nm and 1.2~nm.
The neighbour lists were updated every 5th step with cut-off 1.4~nm. Temperature was coupled separately
for lipids and water to 303~K with the velocity-rescale method~\cite{bussi07} with coupling constant 0.2~ps.
Pressure was semi-isotropically coupled to the atmospheric pressure with the Berendsen method~\cite{berendsen84}.

{\it POPC}
The starting structures for the pure POPC and DOPC simulations was taken from the Slipids~\cite{jambeck12b} website \\ (http://people.su.se/$\sim$jjm/Stockholm\_Lipids/Downloads.html).
The starting structures for mixed POPC/Cholesterol simulations were constructed with the ChARMM-GUI website~\cite{jo08}. 
They contained 100 POPC/24 Cholesterol molecules and 80 POPC/80 Cholesterol molecules for
the simulations of 20\% Cholesterol and 50\% Cholesterol respectively. The TIP3P water model~\cite{jorgensen83} was used to
solvate the system.
The publicly available CHARMM36 forcefield parameters \\ (http://www.gromacs.org/@api/deki/files/184/=charmm36.ff\_4.5.4\_ref.tgz) 
by Piggot et al. \cite{piggot12} were used. Cholesterol parameters came
from Lim et al. \cite{lim12} and were converted into Gromacs files by using the PyTopol tool \\ (https://github.com/resal81/PyTopol).  
Single point energy calculation was done to assess the conversion. 
%(At the time I've made the simulations, the charmm36 ff for cholesterol in gromacs format was not available at
%http://mackerell.umaryland.edu/charmm_ff.shtml#gromacs).
Simulations were performed for 200ns and the last 100ns was used for the calculations. Timestep of 2fs was
used with leap-frog integrator. All bond lengths were constrained with LINCS~\cite{hess97,hess07}. Temperature was maintened at
303 K with the velocity-rescale method~\cite{bussi07} and a time constant of 0.2 ps. Pressure was maintained semiisotropically
at 1 bar using the Parrinello-Rahman algorithm~\cite{parrinello81} with a time constant of 1.0 ps. The neighbour list
was updated every 10th step with a cut-off of 1.2 nm. Lennard-Jones interactions were switched to zero
between 0.8 nm and 1.2 nm. PME~\cite{darden93,essman95} with real space cut-off at 1.2nm was used for electrostatics.

\subsubsection{MacRog}
The lipid force field parameters were obtained from the developers and they correspond to the published DPPC parameters~\cite{maciejewski14} with the inclusion of the 
double bond parameters. This inclusion of unsaturated lipid tails will be published in the near future. 
%\todo{You have recent paper where you use POPC/chol simulations with this model: Kulig et al. BBA 1848 (2015) 422-432 http://dx.doi.org/10.1016/j.bbamem.2014.10.032.
%would this be a correct refence here?}
A bilayer with 288 POPC lipids was hydrated with 12600 TIP3P water~\cite{jorgensen83} molecules ($\sim$44/lipid) and simulated for 100~ns with a time step of 2~fs. Data was saved 
every 10~ps and the first 20~ns of the trajectory was discarded from the analysis. 
All bond lengths were constrained with LINCS~\cite{hess97,hess07}. The temperatures of the lipids and the solvent were separately coupled to the Nos\'{e}--Hoover thermostat~\cite{nose84,hoover85} 
with a target temperature of 310~K and a time constant of 0.4~ps. Semi-isotropical pressure coupling to 1~bar was obtained with the Parrinello--Rahman 
barostat~\cite{parrinello81} with a time constant of 1~ps. PME~\cite{darden93,essman95} was employed to calculate the long-range electrostatic interactions. Lennard-Jones interactions were cut off 
at 1~nm and the dispersion correction was applied to both energy and pressure. A neighbour list with a radius of 1~nm was updated every step. 

Identical parameters were employed for both full hydration and for the dehydration simulations. The dehydration simulations were also run for 100~ns 
with data saved every 10~ps.

The initial structures for the simulations with 10, 40, 50 and 60 mol\% of cholesterol were obtained by replacing 14, 56, 64 or 72 POPC molecules 
with cholesterol molecules in the initial structure containing 128 POPC molecules. These systems were simulated for 400 ns and the first 200 ns was 
discarded from analysis. Data was saved every 100 ps.

\subsubsection{GAFFLipid}
The initial structure in Lipidbook \cite{domanski10} had different glycerol backbone isomers in different leaflets. 
To generate the initial structure we took the structure delivered by Slipid developers~\cite{jambeck12b}. Also this structure
had one lipid with different glycerol bakcbone isomer. This lipid and one lipid from opposite leaflet were removed
after the system was equilibrated.

The force field parameters were generated using files obtained from the Lipidbook website \\ (http://lipidbook.bioch.ox.ac.uk/package/show/id/150.html)~\cite{domanski10}. 
The conversion to Gromacs compatible formats was performed using the acpype tool~\cite{silva12}. The accuracy of the conversion was checked by calculating 
the total energy of a single POPC lipid molecule using the sander program which is part of the AmberTools14 package~\cite{ferrer13} and Gromacs 4.6.5. 
A difference of 0.002 kcal/mol was obtained between the two programs.

Timestep of 2~fs was used in Langevin dynamics with zero friction term and collision frequency of 1.0~ps$^{-1}$. 
Covalent bonds with hydrogens were constrained with LINCS algorithm~\cite{hess97,hess07}.
Coordinates were written every 10~ps. PME~\cite{darden93,essman95} with real space cut-off at 1.0~nm was used 
for electrostatics. Plain cut-off with 1~nm was used for Lennard-Jones interactions. 
The neighbour lists were updated every 5th step with cut-off 1.0~nm. 
Pressure was semi-isotropically coupled to the 1~bar pressure with the Berendsen method~\cite{berendsen84}.
\subsubsection{Lipid14}
%The used simulation files and part of the trajectroy are available~\cite{lipid14files}.
The initial structure was taken directly from the Lipidbook~\cite{domanski10}.
The Amber compatible force field parameters were generated using the tleap program which is integrated in the AmberTools14 package~\cite{ferrer13}. 
A workflow similar to the one used previously for the conversion and validation of the GAFFLipid parameters was followed here. 
As before, a negligible energy difference of 0.003 kcal/mol was obtained between the two programs.

Timestep of 2~fs was used in Langevin dynamics with zero friction term and collision frequency of 1.0~ps$^{-1}$. 
Covalent bonds with hydrogens were constrained with LINCS algorithm~\cite{hess97,hess07}.
Coordinates were written every 10~ps. PME~\cite{darden93,essman95} with real space cut-off at 1.0~nm was used 
for electrostatics. Plain cut-off with 1~nm was used for Lennart-Jones interactions. Dispersion correction
was used for energy and pressure. The neighbour lists were updated every 5th step with cut-off 1.0~nm. 
Pressure was semi-isotropically coupled to the 1~bar pressure with the Berendsen method~\cite{berendsen84}.

\subsubsection{Poger et al.}
The Poger lipids are derived from GROMOS G53A6~\cite{poger10} and were initially coined 53A6-L (L for lipids), and are now part of GROMOS G54A7~\cite{poger12}. 
They work with the SPC water model~\cite{berendsen81}. The initial hydrated bilayer structure of 128 DPPC/5841 water molecules as well as force field parameters were downloaded 
from David Poger's web site \\ (http://compbio.chemistry.uq.edu.au/~david/) on April 2012. 
We noticed that the same files downloaded in October 2013 appear to lack two dihedral angles in the choline headgroup (only one dihedral of type gd\_29 allowing 
the rotation of the 3 choline methyls) compared to the April 2012 version (3 dihedrals of type gd\_29 for the 3 choline methyls). This should not affect the 
bilayer structure and only change the kinetics of the choline methyls rotation, however the October 2013 version has not been tested.

MD Simulations (two repetitions with independent initial velocities) were run for 100 ns using a 2 fs time step and the analysis 
was performed on the last 50 ns. Coordinates were saved every 50 ps for analysis. All bond lengths were constrained with the LINCS algorithm~\cite{hess97,hess07}. Temperature was kept 
at 323K with the v-rescale~\cite{bussi07} thermostat with a time constant of 0.1 ps (DPPC and water coupled separetly). Pressure was maintained semi-isotropically at 1 bar using 
the Parrinello-Rahman barostat~\cite{parrinello81} using a 4 ps time constant and a compressibility of 4.5e-5 bar$^{-1}$. For non-bonded interactions, two conditions were tested:
i) A 0.8-1.4 nm twin-range cutoff with the neighbor list updated every 5 steps for both electrostatics and Lennard-Jones. For the former the generalized reaction 
field (RF) with a dielectric permitivity of 62 was used beyond the 1.4 nm cutoff~\cite{tironi95}. This is the original setup that Poger et al.~\cite{poger10} used.
ii) PME~\cite{darden93,essman95} electrostatics with a real space cutoff of 1.0 nm, a Fourier spacing of 0.12 nm and an interpolation order of 4, LJ computed with a 1.0-1.4 nm twin-range cutoff, 
neighbor list updated every 5 steps. Note that Poger and Mark tested the effect of PME vs RF in the ref.~\cite{poger12}, but used a 1.0 nm cutoff with PME and 1.4 nm with RF for LJ 
interactions. Since 0.8-1.4 nm twin-range cutoff for LJ interations is used in the parametrization of GROMOS force field we decided to use that
also in the simulations with PME.
%Since there is no dispersion correction in GROMOS (and since dispersion correction should not be used anyway for systems with interfaces), we decided to stick 
%to the same LJ cutoff as that with the RF scheme i), that is 1.4 nm.

Since Poger lipids come from GROMOS force field, it is important to note that GROMOS uses the RF scheme for computing electrostatics (this is the method used for the 
force field parameterization). Using setup i) based on RF, we were able to reproduce the results (i.e. area per lipid~0.63 nm$^2$) from the original work only with 
GROMACS version $<=$ 4.0.* (the original authors~\cite{poger10} used GROMACS version 3.3.3). On going to versions $>=$ 4.5.*, the area per lipid dropped below 0.58 nm$^2$. 
The GROMACS developers were contacted and a redmine issue opened (http://redmine.gromacs.org/issues/1400). The difference comes from the new Trotter decomposition 
introduced in version 4.5. A fix has been introduced in version 4.6.6 that allows a recovery of ~0.615 nm$^2$. The results in terms of area per lipid using the different 
GROMACS versions are here~\cite{fuchs14}.
% http://www.dsimb.inserm.fr/~fuchs/project\_Samuli/Poger\_DPPC/tests\_gmx\_versions/)[Samuli, you may want to add this Figure+Legend+Comment 
%o a stable link (Figshare?), let me know]. 
%Of note, some tests (Miguel Machuqueiro, personnal communication) with the same RF scheme i) on simulations with proteins 
%(thus with explicit hydrogens) showed that this setup is still unstable and eventually leads to crashes with versions 4.6.6 and 4.6.7 (and so does version 4.5.*).
Thus we decided to use only the PME setup ii) for computing the order parameter since it gives stable results whatever the GROMACS version. We obtained an area per 
lipid of ~0.615 nm$^2$, below 0.648 nm$^2$ found by the original authors with their PME setup (see~\cite{poger12}). We explained that by the fact we used 
a 1.4 nm for the LJ cutoff and they used 1.0 nm. 

\subsubsection{Slipid}
%The used simulation files and part of the trajectroy are available~\cite{slipidsFILES}.

Initial coordinates for a hydrated DPPC bilayer (30 waters/lipid) at 323K were taken directly
from \\ http://people.su.se/\~jjm/Stockholm\_Lipids/Downloads.html  The Slipids force field~\cite{jambeck12} was used for the the all atom description of DPPC, and
water was described with the TIP3P water model~\cite{jorgensen83}. Simulations were performed within the NPT ensemble using the GROMACS 4.6.1 simulation
package~\cite{hess08}. The nose--hoover rescaling thermostat~\cite{nose84,hoover85} was used with reference temperature of
323 K and a relaxation time constant of 0.5 ps. Water and lipids were coupled separately to
the heat bath. Pressure was kept constant at 1.013 bar using a semi--isotropic Parinello--Rahman
barostat~\cite{parrinello81} with a time constant of 10.0 ps. Equations of motion were
integrated with the leapfrog algorithm using a timestep of 2 fs. Long range
electrostatic interactions were calculated using the PME method~\cite{darden93,essman95}, with a fourth order
smoothing spline. A real space cutoff of 1.0 nm was employed with grid spacing of 0.12 in the reciprocal space.
Lennard--Jones potentials were cutoff at 1.4 nm, with dispersion correction applied both to
energy and pressure. All covalent bonds in lipids were constrained using the LINCS algorithm~\cite{hess97}, 
whereas water molecules were constrained using SETTLE~\cite{miyamoto92}. Twinrange cutoffs,
1.0 nm and 1.6 nm, were used for the neighborlists with the longrange neighbor list updated every
10 steps. This simulation protocol corresponds to the protocol used in Ref~\cite{jambeck13}. 
%The simulations were run for XXX ns with the last XXX ns used for calculating the order parameters. 
\subsubsection{Kukol}
A bilayer patch with 512 POPC lipids was constructed and hydrated with $\sim$40 SPC water molecules per lipid. 
The force field parameters were obtained from Lipidbook \cite{domanski10}.
This bilayer was simulated with a 2~fs time step for a total of 50~ns and coordinates were saved every 100~ps. 
All bonds were constrained with LINCS~\cite{hess97,hess07}. PME~\cite{darden93,essman95} was employed for the long-range electrostatics. Lennard-Jones interactions 
were cut off at 1.4~nm. A neighbour list with a radius of 0.8~nm was updated every 5~steps. The constant temperature of 298~K 
was maintained with the Berendsen thermostat \cite{berendsen84} with a time constant of 0.1~ps. The Berendsen barostat \cite{berendsen84} 
was employed for semi-isotropical pressure coupling at 1~bar.

%Additionally, the reaction field electrostatics with a twin-range cut-off scheme [cite GROMOS] was employed yet the results 
%were consistent with those obtained with PME. (This paragraph can be discarded.)

\subsubsection{Chiu et al.}
The force field parameters and the initial configuration were available through the Lipidbook~\cite{domanski10}.
Timestep of 2~fs was used with leap-frog integrator. Covalent bond lengths were constrained with LINCS algorithm~\cite{hess97,hess07}. 
Coordinates were written every 10~ps. PME~\cite{darden93,essman95} with real space cut-off at 1.0~nm was used 
for electrostatics. Twin range cut-off was used for the Lennardt-Jones interactions with short and long cut-offs at 1.0~nm and 1.6~nm, respectively.
The neighbour lists were updated every 5th step with cut-off at 1.0~nm. Temperature was coupled separately
for lipids and water to 298~K with the velocity-rescale method~\cite{bussi07} with coupling constant 0.2~ps.
Pressure was semi-isotropically coupled to the atmospheric pressure with the Parrinello-Rahman method~\cite{parrinello81}.
%\subsubsection{H\"ogberg et al. and Rabonovich et al.}
%Simulations done in \cite{hogberg08} and \cite{rabinovich14} are used.

%Time step 2fs, Ewald summation,
%cut-off 14 Å, update of neighbour list each 10 steps,
%Nose-Hoover thermostat,  Parrinello-Rahman barostat,
%semi-anizotropic pressure coupling,
%long-range isotropic pressure correction
\subsubsection{Ulmschneider}
The initial structure containing 128 POPC molecules with 3328 TIP3P water~\cite{jorgensen83} molecules (26 per lipid) was downloaded from Lipidbook \cite{domanski10} 
together with the topologies. This bilayer was simulated for 100~ns with a time step of 2~fs and the data was saved every 10~ps. The bonds involving hydrogen atoms were 
constrained with LINCS~\cite{hess97,hess07}. The temperature was kept at 298~K with the Berendsen thermostat~\cite{berendsen84}. The pressure was semi-isotropically coupled to the Berendsen 
barostat~\cite{berendsen84} with a time constant of 1~ps and a target pressure of 1~bar. PME~\cite{darden93,essman95} was employed for long range electrostatics and a cut-off of 1~ns was employed for 
the Lennard-Jones interactions. A neighbour list with a radius of 1~nm was updated every 10~steps. 

Additionally, the simulations were repeated with the dispersion correction applied to pressure and temperature. Even though the area per lipid decrease
d slightly, the head group order parameters were only slightly affected.

\subsubsection{Tj\"ornhammar et al.}
%The used simulation files and part of the trajectroy are available~\cite{tjornhammarfiles}.

The gel phase structure delivered by Tj\"ornhammar  and edholm~\cite{tjornhammar14} was ran a 5ns at 70 degrees in order to destroy the 
ordered gel configuration. This was followed by 200ns simulation at 50 degrees. The last 100ns of this simulation was used for analysis. 
The same mdp file as in the paper's~\cite{tjornhammar14} SI was used except for the temperatures.

\subsubsection{CHARMM36-UA}
A hydrated bilayer consisting of 128 DLPC lipids and 3840 water molecules is modeled by the force field of Lee and co-workers~\cite{lee14}, 
which is a combination of the all-atom CHARMM36 force-field~\cite{klauda10} and the united-atom Berger model~\cite{berger97}. 
The nonbonding interactions are calculated using an atom-based switching function with inner and outer cutoffs of 0.8 and 1.2 nm~\cite{lee14}. 
Long range electrostatic interactions are implemented using the particle-particle particle-mesh solver with a relative accuracy of 0.0001. The system 
is first equilibrated for 30 ns in the NP$\gamma$T ensemble (Nose-Hoover~\cite{nose84,hoover85} style thermostat and barostat with anisotropic pressure coupling) 
at 323 K and 1 bar with timestep of 1 fs, the next 20 ns of dynamics are taken for calculation of configurational averages. 
%\todo{The simulation times given currently contradict with the Table~\ref{systems}.}
Simulations were 
carried out by using the LAMMPS package~\cite{plimpton95}. %( http://doi.org/10.1006/jcph.1995.1039 )., the input files are available ( http://dx.doi.org/10.5281/zenodo.13821 ).

%\subsubsection{OPLS-AA}
%\todo{Joona Tynkkynen. We have this only with 150mM of NaCl delivered by Joona Tynkkynen. Options are to remove these results from this publicatio, run the simulations without ions (if not yet available), 
%  or include the results with ions. In my understanding this is a proto version of MacRog so it could be left out as well. However, for historical reasons and 
%to understand literature it might be useful to include also these results. COMMENT: Matti Javanainen suggested that we would remove these results. If there are no objections and the
%data without ions will not be delivered, we will remove this.}
\onecolumngrid
\subsection{Dihedral angle distributions as a function of cholesterol in CHARMM36}
\begin{figure}[!h]
  \centering
  \includegraphics[width=17.2cm]{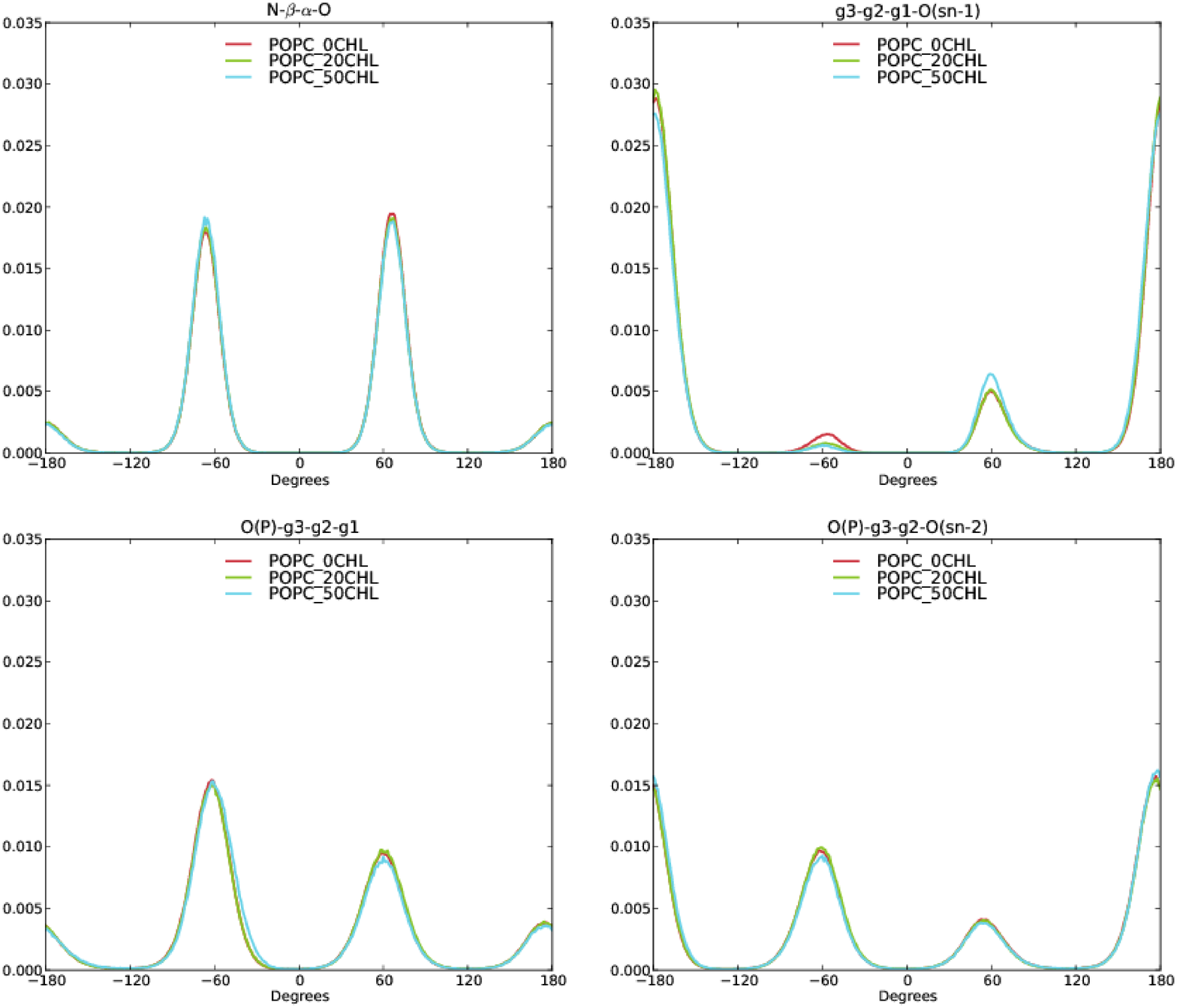}
  \caption{\label{dihsCHOLcharmm}
    The effect of cholesterol content on the glycerol backbone and choline dihedral angles in CHARMM36 model.}
\end{figure}

%\newpage

%\onecolumngrid

% \listoftodos

\end{document}